\documentclass[referee]{raa}
\usepackage{graphicx,times}

\begin{document}

   \title{Similarity of Jet Radiation between Flat Spectrum Radio Quasars and GeV Narrow-Line Seyfert 1 Galaxies: A Universal $\delta$--$L_{\rm c}$ Correlation}

   \volnopage{Vol.0 (200x) No.0, 000--000}      
   \setcounter{page}{1}          

   \author{Yong-Kai Zhu
      \inst{1}
   \and Jin Zhang
      \inst{2,4}
   \and Hai-Ming Zhang
      \inst{1}
   \and En-Wei Liang
	 \inst{1,2}
   \and Da-Hai Yan
	 \inst{3}
   \and  Wei Cui
	 \inst{4}
   \and Shuang-Nan Zhang
	 \inst{2,3}
   }

   \institute{Guangxi Key Laboratory for Relativistic Astrophysics, Department of Physics, Guangxi University,
		  Nanning 530004, China;\\
        \and
             Key Laboratory of Space Astronomy and Technology, National Astronomical Observatories, Chinese Academy of Sciences, Beijing 100012, China; {\it jinzhang@bao.ac.cn}\\
        \and
             Key Laboratory of Particle Astrophysics, Institute of High Energy Physics, Chinese Academy of Sciences,
			Beijing 100049, China;\\
	   \and
		  Department of Physics and Astronomy, Purdue University, West Lafayette, IN 47907, USA\\
   }

   \date{Received~~2009 month day; accepted~~2009~~month day}

\abstract{By modeling the broadband spectral energy distributions (SEDs) of a typical flat spectrum radio quasar (FSRQ, 3C 279) and two GeV narrow-line Seyfert 1 galaxies (NLS1s, PMN J0948+0022 and 1H 0323+342) in different flux stages with the one-zone leptonic models, we find a universal correlation between their Doppler factors ($\delta$) and peak luminosities ($L_{\rm c}$) of external Compton scattering bumps. Compiling a combined sample of FSRQs and GeV NLS1s, it is found that both FSRQs and GeV NLS1s in different stages and in different sources well follow the same $\delta$--$L_{\rm c}$ correlation. This indicates that the variations of observed luminosities may be essentially due to the Doppler boosting effect. And the universal $\delta$--$L_{\rm c}$ relation between FSRQs and GeV NLS1s in different stages may be further evidence that the particle acceleration and radiation mechanisms for the two kinds of sources are similar. In addition, by replacing $L_{\rm c}$ with the observed luminosity in the \emph{Fermi}/LAT band ($L_{\rm LAT}$), this correlation holds, and it may serve as an empirical indicator of $\delta$. We estimate the $\delta$ values with $L_{\rm LAT}$ for 484 FSRQs in the \emph{Fermi}/LAT Catalog and they range from 3 to 41, with a median of 16, which are statistically consistent with the values derived by other methods.
\keywords{galaxies: active---galaxies: jets---quasars: general---galaxies: Seyfert}
}

   \authorrunning{Y. K. Zhu \& J. Zhang et al. }            
   \titlerunning{Similarity of Jet Radiation between FSRQs and GeV NLS1s}  
A

   \maketitle

%
%

\section{Introduction}           
Flat spectrum radio quasars (FSRQs) and BL Lacertae objects (BL Lacs) are referred to as blazars, whose broadband spectral energy distributions (SEDs) are thought to be dominated by the jet emission. FSRQs are different from BL Lacs for having significant emission lines. It was proposed that many radio-loud (RL) narrow-line Seyfert 1 galaxies (NLS1s) display blazar characteristics and maybe also host relativistic jets\footnote{Gu et al. (2015) studied the compact radio structures of 14 NLS1s with Very Long Baseline Array observations at 5 GHz and reported that 50\% of the sources show a compact core only and the remaining 50\% exhibit a core-jet structure.} (Zhou et al. 2003; Yuan et al. 2008); this has been confirmed by the detection of $\gamma$-ray emission from NLS1s by \emph{Fermi}/LAT (Abdo et al. 2009; D'Ammando et al. 2012) and the observations of Kiloparsec--parsec scale radio structures (Doi et al. 2012; Gu et al. 2015), especially the observation of apparent superluminal velocity in the jet of SBS 0846+513. The broadband SEDs of GeV NLS1s are similar to blazars (Abdo et al. 2009; Zhang et al. 2013b; Paliya et al. 2013; Sun et al. 2014, 2015; Kreikenbohm et al. 2016; Paliya \& Stalin 2016) and their $\gamma$-ray emission is also dominated by external Compton scattering (EC) process by photons from their broad-line regions (BLRs) (e.g., Sikora et al. 1994; Ghisellini et al. 2009; Zhang et al. 2014; Chen et al. 2012; Liao et al. 2014; Sun et al. 2015). GeV NLS1s have significant emission lines, and sometimes a blue bump of the disk thermal radiation is observed in their SEDs. Therefore, the circumnuclear environment in NLS1s is analogous to that in FSRQs. Recently, on the basis of one-zone leptonic jet models, Sun et al. (2015) reported that the jet property of GeV NLS1s, including their jet power, radiation efficiency, and magnetization parameter, is indeed a bridge between FSRQs and BL Lacs, but more analogous to FSRQs than BL Lacs. Further more, Zhang et al. (2015) suggested a BL Lac--NLS1--FSRQ sequence with the increase of their BLR luminosity and Eddington ratio, which may correspond to the change of the accretion disk structure and the transformation of the dominant mechanism for jet launching.

The luminosities of blazars are thought to be boosted since the emitting regions move with relativistic velocity and small viewing angle ($\theta$). Recently, Richards \& Lister (2015) reported that the jets of RL NLS1s are aligned at moderately small angles to the line of sight, which is similar to blazars. The active galactic nuclei (AGNs) that are not detected with the \emph{Fermi}/LAT have, on average, lower Doppler factors than those that are detected with the \emph{Fermi}/LAT (Lister et al. 2015). The measurements of these parameter values are very crucial for understanding the physics of jets (e.g., Nokhrina et al. 2015), for examining the unified models of AGNs (e.g., Hovatta et al. 2009; Savolainen et al. 2010), and even for investigating the intrinsic radiation physics of blazars with gamma-ray bursts (Wu et al. 2011; Wang et al. 2011;  Nemmen et al. 2012; Zhang et al. 2013a). Some approaches have been proposed to estimate the Doppler factor values of AGNs. With the Very Long Baseline Interferometry (VLBI) measurements of the core angular size and radio flux, Ghisellini et al. (1993) estimated the $\delta$ values of $\sim$100 AGNs by comparing the observed X-ray fluxes to that predicted by the Self-Synchrotron-Compton scattering model. The derived $\delta$ values with this method usually have large uncertainty since it needs simultaneous X-ray and VLBI observations and strongly depends on the turnover values (L\"{a}hteenm\"{a}ki \& Valtaoja 1999). Jorstad et al. (2005) also used the VLBI observation data to derive the Doppler factors by comparing the flux decline timescale ($\tau_{\rm obs}$) with the light-travel time ($\tau_{\rm int}$) across the emitting region ($\tau_{\rm obs}\sim\tau_{\rm int}\delta$). Another more popular way to estimate the Doppler factors is to obtain the variability brightness temperatures of sources using total flux density variations (L\"{a}hteenm\"{a}ki \& Valtaoja 1999; Hovatta et al. 2009), which is boosted by $\delta^3$ compared with the intrinsic brightness temperature of the source. Although the superluminal motions in many of sources were resolved by VLBI observations (Homan et al. 2001; Kellermann et al. 2004; Jorstad et al. 2005; Piner et al. 2007; Lister et al. 2013), a quantitative assessment of the beaming parameters (i.e., the bulk Lorentz factor $\Gamma$ or the velocity of emitting region and the viewing angle $\theta$) is still lacking. With the transparency condition, one may also estimate the lower limit of $\delta$ (e.g., Fan et al. 2014). Theoretically, by modeling the observed broadband SEDs, the Doppler factors can also be constrained (Zhang et al. 2012, 2014, 2015; Sun et al. 2015; Yan et al. 2014; Kang et al. 2014).

It is interesting that a tentative correlation between $\delta$ and the peak luminosities of the EC bumps ($L_{\rm c}$) is found for FSRQ 3C 279 (Zhang et al. 2013c) with four SEDs and two GeV NLS1s (Sun et al. 2015). In this paper, we firstly further test this correlation in individual sources with 14 SEDs of 3C 279. And then we compile a sample of FSRQs and GeV NLS1s to study the correlation between $L_{\rm c}$ and $\delta$ for different sources and also investigate whether this correlation can be used to estimate $\delta$ with the observed luminosity. The analogous observations in both FSRQs and GeV NLS1s also motivate us to explore whether they share the same $\delta$--$L_{\rm c}$ relation and the physics of this correlation. Model and SED fitting and the $\delta$--$L_{\rm c}$ correlation in different stages for 3C 279 are presented in \S 2. The $\delta$--$L_{\rm c}$ correlation in different sources is described in \S 3. The possible physical implications of this correlation are discussed in \S 4. Using this relation to derive the Doppler factors of FSRQs in \emph{Fermi}/LAT Third Source Catalog (3FGL) and comparing the results with others are reported in \S 5. Summary and conclusions are given in \S 6.

\section{The $\delta$--$L_{\rm c}$ Correlation for 3C 279}

The 14 broadband SEDs observed in different stages for 3C 279 are collected from literature (Hayashida et al. 2012, 2015; Paliya et al. 2015) and shown in Figure \ref{SED}. Following our previous works (Zhang et al. 2014, 2015; Sun et al. 2015), the simple one-zone leptonic model is used to explain the 14 observed broadband SEDs of 3C 279 in different stages. Some authors suggested that the seed photons from torus may present the better fits to the $\gamma$-ray spectra than those from BLRs (Sikora et al. 2009; Tavecchio \& Mazin 2009; Tavecchio et al. 2013) due to the Klein--Nashina (KN) effect in the TeV band and the $\gamma$-ray attenuation (via pair production) at energies above 10 GeV (Liu \& Bai 2006). However, some observations also indicate that the $\gamma$-ray emitting regions of blazars should be inside the BLRs. For example, Poutanen \& Stern (2010) reported that the spectral breaks in the 2--10 GeV range can be well reproduced by the absorption of $\gamma$-rays via photon--photon pair production on the He~{\scriptsize II} Lyman recombination continuum and lines; Le\'{o}n-Tavares et al. (2013) and later Isler et al. (2013) also reported that the correlation of the increased emission line flux with millimeter core ejections and $\gamma$-ray, optical, and ultraviolet flares implies that the BLR extends beyond the $\gamma$-emitting region during the 2009 and 2010 flares for 3C 454.3. Considering the consistency of SED fitting models with our previous works, therefore, the single-zone synchrotron+IC model is still used to explain the jet emission of 3C 279, where the IC process includes both the SSC process and the external Compton scattering of BLR photons (EC/BLR). The KN effect and the absorption of high energy $\gamma$-ray photons by extragalactic background light (Franceschini et al. 2008) are also taken into account.

In the SED fitting, the radiation region is assumed to be a sphere with radius $R$. The electron distribution is taken as a broken power law, which is characterized by an electron density parameter ($N_0$, in units of cm$^{-3}$), a break energy $\gamma_{\rm b}$ and indices ($p_1$ and $p_2$) in the range of  $\gamma_{\rm e}\sim[\gamma_{\min}, \gamma_{\max}]$. The energy density of BLR for 3C 279 is derived with the BLR luminosity (Zhang et al. 2014, 2015) and is taken as $U^{'}_{\rm BLR}=3.91\times10^{-2}\Gamma^2$ erg cm$^{-3}$ (Table 1 in Zhang et al. 2014) in the jet comoving frame. Since for blazars we are likely looking at the jet within the $1/\Gamma$ ($\Gamma$ is the bulk Lorentz factor) cone, and that the probability is the highest at the rim of the cone, we take $\delta=\Gamma$ in all the calculations, i.e., the viewing angle is equal to the opening angle of the jet\footnote{It is well known that there is a
Doppler boosting effect in the radiation of blazars. If the viewing angle is larger than the opening angle of the jet, the leptonic models would not be able to explain the observation data as reported in Zhang et al. (2015).}. Adding the magnetic field strength ($B$) of the radiating region, the model can be described with nine parameters: $R$, $B$, $\delta$, and the electron spectrum parameters ($\gamma_{\min}$, $\gamma_{\rm b}$, $\gamma_{\max}$, $N_0$, $p_1$, $p_2$). We take $R=\delta c\Delta t/(1+z)$, where $z=0.536$ is the redshift of 3C 279 and $\Delta t$ is the variability timescale and the $\Delta t$ values are listed in Table 1. The indices of $p_1$ and $p_2$ are derived from the spectral indices of the observed SEDs as reported by Zhang et al. (2012). $\gamma_{\max}$ is usually poorly constrained, but it does not significantly affect our results and is fixed at $\gamma_{\max}=5000$. Hence the free parameter set of our SED modeling is $\left\{B,\delta, N_{0}, \gamma_{\rm b}, \gamma_{\rm min}\right\}$. Following our previous works (Zhang et al. 2014, 2015; Sun et al. 2015), the $\chi^2$ minimization technique is also used to perform the SED fits. For the details of this technique and fitting strategies please refer to Zhang et al. (2014, 2015) and Sun et al. (2015). The SED fits are shown in Figure \ref{SED}, and the model parameters are reported in Table 1. As shown in Figure \ref{SED}, the significant variabilities of luminosity for 3C 279 can be observed.

\begin{figure}
   \centering
   \includegraphics[width=\textwidth, angle=0]{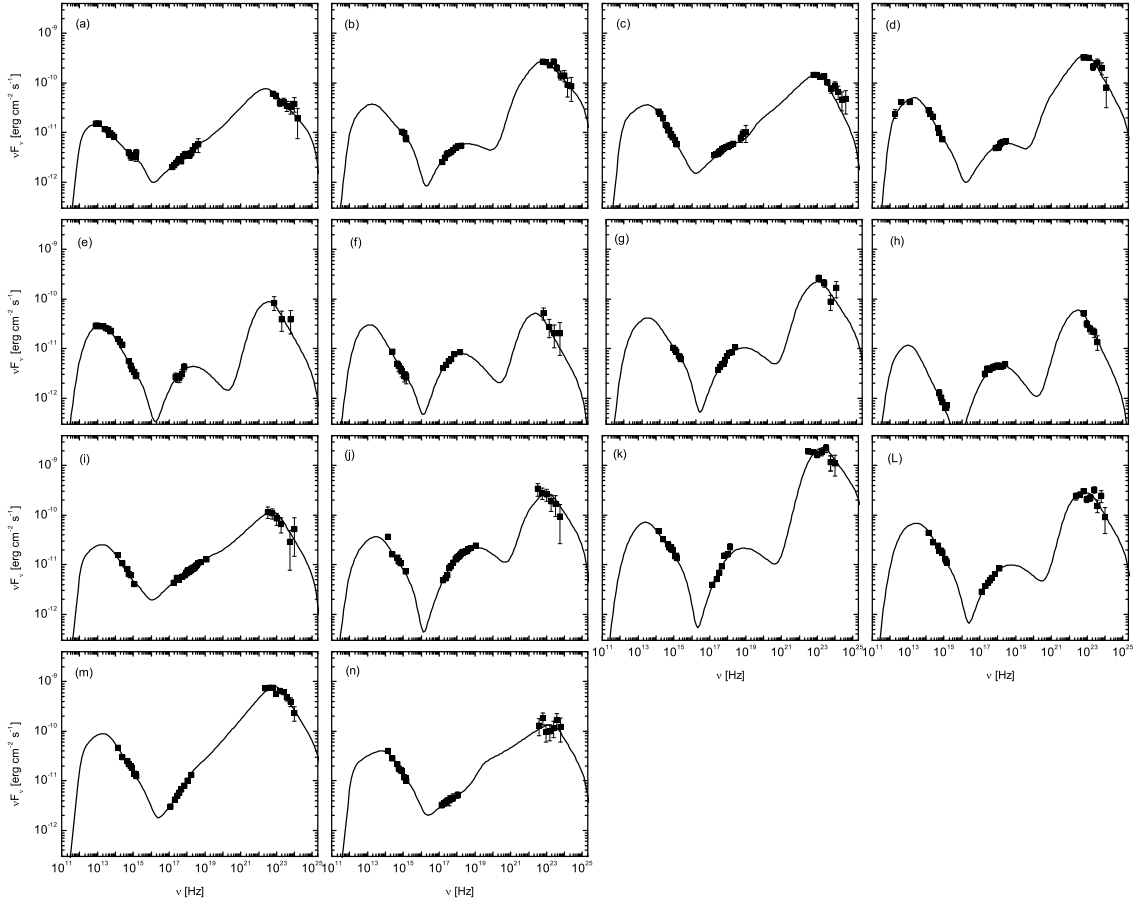}
\caption{Observed SEDs (scattered data points) with model fitting (lines) for 3C 279. The observation data of panels (a)--(h), panels (i)--(k), and panels (L)--(n) are from Hayashida et al. (2012), Hayashida et al. (2015), and Paliya et al. (2015), respectively.}\label{SED}
   \label{SED}
\end{figure}

\begin{table}
\begin{center}
	\caption{Derived Parameters for 3C 279 with the One-zone Leptonic Model}
	\label{tab:1}
	\begin{tabular}{lcccccccccc} 
		\hline\noalign{\smallskip}
		SEDs$^{a}$  &  $B$   &$\delta$ & $\Delta t$$^{b}$  & $p_1$ &  $p_2$ &  $\gamma_{\min}$ & $\gamma_{\rm b}$ & $\log N_{0}$ & $\log L_{\rm c}$ & $\log L_{\rm LAT}^{c}$\\
 		& [G]  &  & [hr] &   & &  &  & [cm$^{-3}$]&  [erg s$^{-1}$]&  [erg s$^{-1}$]\\

		\hline\noalign{\smallskip}
			a&4.5$\pm$0.7&12.2$\pm$1.0&12&2.2&3.76&$1^{+2}_{-0}$&214$\pm$123&5.35$\pm$0.12&46.90$\pm$0.16&47.41\\
			b&4.5$\pm$0.5&14.7$\pm$0.7&12&1.9&3.82&53$\pm$21&288$\pm$60&4.40$\pm$0.10&47.47$\pm$0.08&48.14\\
			c&5.6$\pm$0.5&13.4$\pm$0.3&12&2.28&4.38&15$\pm$6&447$\pm$41&5.26$\pm$0.05&47.20$\pm$0.07&47.87\\
			d&4.8$\pm$0.8&15.3$\pm$0.8&12&1.9&4.28&$1^{+99}_{-0}$&324$\pm$60&4.27$\pm$0.10&47.60$\pm$0.16&48.18\\
			e&6.7$\pm$1.9&12.7$\pm$1.8&12&1.9&4.28&$1^{+112}_{-0}$&275$\pm$63&4.53$\pm$0.23&46.91$\pm$0.31&47.48\\
			f&6.6$\pm$1.6&10.7$\pm$1.2&12&1.9&4.28&83$\pm$10&295$\pm$82&4.95$\pm$0.11&46.72$\pm$0.32&47.21\\
			g&4.7$\pm$0.8&13.1$\pm$1.1&12&1.9&4.08&81$\pm$15&355$\pm$90&4.62$\pm$0.11&47.36$\pm$0.17&48.03\\
			h&3.9$\pm$0.8&10.5$\pm$1.3&12&1.5&4.48&$1^{+114}_{-0}$&251$\pm$69&4.20$\pm$0.11&46.80$\pm$0.20&47.15\\
			i&4.9$\pm$0.9&12.0$\pm$1.0&12&2.3&4.2&3$\pm$1&389$\pm$107&5.63$\pm$0.10&47.05$\pm$0.14&47.59\\
			j&3.9$\pm$0.7&12.3$\pm$1.3&12&1.9&4.1&94$\pm$11&468$\pm$75&4.87$\pm$0.10&47.47$\pm$0.30&48.04\\
			k&3.3$\pm$0.8&20.5$\pm$1.9&6&1.5&4.0&102$\pm$28&339$\pm$101&3.82$\pm$0.12&48.34$\pm$0.15&49.06\\
			L&6.2$\pm$1.1&14.9$\pm$1.1&12&2.4&4.24&87$\pm$14&389$\pm$107&5.49$\pm$0.12&47.47$\pm$0.12&48.22\\
			m&5.3$\pm$0.7&18.6$\pm$0.8&12&2.12&4.18&1$\pm$0&275$\pm$63&4.47$\pm$0.07&47.90$\pm$0.06&48.55\\
			n&7.2$\pm$1.6&14.3$\pm$1.5&12&2.48&4.2&$1^{+54}_{-0}$&479$\pm$276&5.55$\pm$0.25&47.14$\pm$0.24&47.95\\
		\noalign{\smallskip}\hline
	\end{tabular}
\\
\end{center}
$^{\rm a}${The SEDs are corresponding to the panel names in Figure \ref{SED}.}\\
$^{\rm b}${Since the $\gamma$-ray spectra for 3C 279 are extracted from some periods and are the averaging fluxes over larger time intervals (Hayashida et al. 2012, 2015; Paliya et al. 2015), $\Delta t=12$ hr is taken following our previous works (Zhang et al. 2012, 2014), except for one observed SED, which is extracted at the peak of flare of $\sim6$ hr (Hayashida et al. 2015). }
$^{\rm c}${The observed luminosity in the LAT band, from 100 MeV to 100 GeV.}
\end{table}

\begin{figure*} 
   \centering
   \includegraphics[width=\textwidth, angle=0]{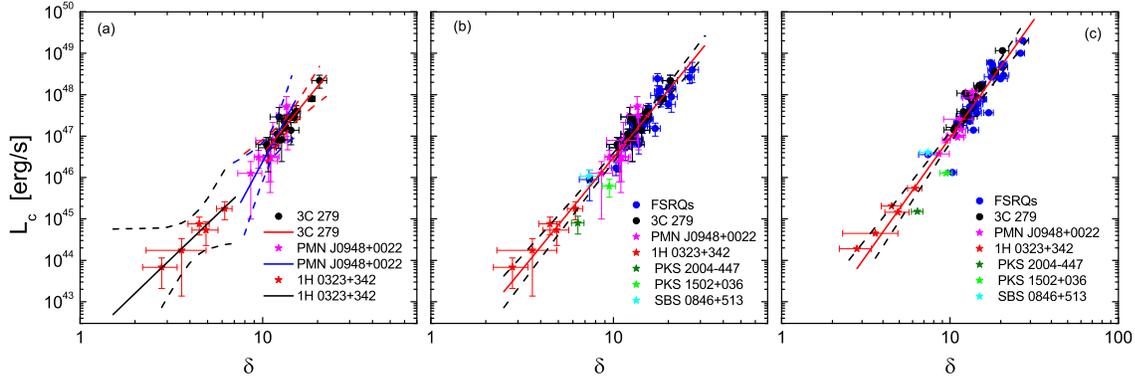}
 \caption{$L_{\rm c}$ (\emph{Panels a, b}) and $L_{\rm LAT}$ (\emph{Panel c}) as a function of $\delta$. \emph{Panel (a)}---The solid lines are the bisectors of the two OLS lines of linear regression fits for the three sources separately. The dashed lines indicate the 3$\sigma$ confidence bands for the linear fits. \emph{Panels (b) (c)}---The solid line is the linear fitting result for the combined FSRQ and GeV NLS1 sample (all the data points). The dashed lines indicate the 3$\sigma$ confidence bands of the linear fit. \emph{Panel (c)}---Replacing $L_{\rm c}$ with the observed luminosity in the \emph{Fermi}/LAT band ($L_{\rm LAT}$) in the $\delta$--$L_{\rm c}$ plane. The parameters of all the fitting lines are given in Table 2.}
\label{Lc-delta}
\end{figure*}

The data of 14 SED fits for FSRQ 3C 279 are presented in the $\delta$--$L_{\rm c}$ plane, as shown in Figure \ref{Lc-delta}(a). Note that the linear fitting results depend on the specification of dependent and independent variables, especially when the data have large error bars or large scatter (Isobe et al. 1990). Therefore three methods of linear regression fits are taken into account here, i.e., ordinary least-squares (OLS) regression of \emph{Y} on \emph{X}, OLS regression of \emph{X} on \emph{Y}, and the bisector of the two OLS lines. The linear fitting results together with the results of the Pearson correlation analysis are reported in Table 2. In order to avoid specifying independent and dependent variables, the bisector of the two OLS lines of linear regression fit is used in the following analysis. The linear fits yield $\log L_{\rm c}/{\rm erg\ s}^{-1}=(41.11\pm0.39)+(5.45\pm0.33)\log\delta$ with a Pearson correlation coefficient of $r=0.93$ and chance probability of $p=1.8\times10^{-6}$, indicating that $L_{\rm c}$ is tightly correlated with $\delta$ for 3C 279 in different stages.

\begin{table}
\begin{center}
\caption{Results of the Pearson Correlation Analysis and the Linear Regression Fits with Three Methods for the Parameter Sets ($X$--$Y$). }
\label{tab:3}
\setlength{\tabcolsep}{1.6pt}
\small
	\begin{tabular}{lcccccccc}

	\hline\noalign{\tiny}
	   &\multicolumn{2}{c}{OLS($X|Y$)}&\multicolumn{2}{c}{OLS($Y|X$)}&\multicolumn{2}{c}{OLS bisector}&\multicolumn{2}{c}{Correlation}\\\cline{2-3}\cline{4-5}\cline{6-7}\cline{8-9}\\
 & $a$ & $b$ & $a$ & $b$ & $a$ & $b$ & $r$ & $p$  \\
	\hline\noalign{\tiny}

$\delta$--$L_{\rm c}$(3C 279)&41.55$\pm$0.49&5.06$\pm$0.42&40.61$\pm$0.50&5.90$\pm$0.43&41.11$\pm$0.39&5.45$\pm$0.33&0.93&$1.8\times10^{-6}$\\
$\delta$--$L_{\rm c}$(30 FSRQs)&41.40$\pm$0.33&5.10$\pm$0.29&40.47$\pm$0.59&5.89$\pm$0.51&40.97$\pm$0.43&5.46$\pm$0.37&0.93&$9.3\times10^{-14}$\\
$\delta$--$L_{\rm c}$(All FSRQs)&41.55$\pm$0.31&5.00$\pm$0.27&40.59$\pm$0.49&5.82$\pm$0.42&41.10$\pm$0.36&5.38$\pm$0.31&0.93&$\sim0$\\
$\delta$--$L_{\rm c}$(1H 0323+342)&41.99$\pm$0.07&4.14$\pm$0.14&41.88$\pm$0.12&4.32$\pm$0.24&41.93$\pm$0.08&4.23$\pm$0.18&0.98&$3.7\times10^{-3}$\\
$\delta$--$L_{\rm c}$(PMN J0948+002)&38.97$\pm$1.19&7.45$\pm$1.14&37.26$\pm$1.47&9.07$\pm$1.36&38.20$\pm$1.27&8.18$\pm$1.19&0.91&$7.7\times10^{-4}$\\
$\delta$--$L_{\rm c}$(All NLS1s)&41.26$\pm$0.28&5.19$\pm$0.32&40.91$\pm$0.39&5.59$\pm$0.42&41.09$\pm$0.33&5.39$\pm$0.36&0.96&$5.5\times10^{-10}$\\
$\delta$--$L_{\rm c}$(All) & 41.25$\pm$0.18 & 5.24$\pm$0.16 & 40.90$\pm$0.25 & 5.57$\pm$0.23 & 41.08$\pm$0.21 & 5.40$\pm$0.19 &0.97&$\sim0$\\
$\delta$--$L_{\rm LAT}$(3C 279)&41.02$\pm$0.50&6.07$\pm$0.43&40.22$\pm$0.48&6.77$\pm$0.43&40.64$\pm$0.42&6.40$\pm$0.37&0.95&$2.9\times10^{-7}$\\
$\delta$--$L_{\rm LAT}$(30 FSRQs)&41.48$\pm$0.61&5.45$\pm$0.50&40.02$\pm$1.00&6.69$\pm$0.82&40.82$\pm$0.75&6.01$\pm$0.62&0.90&$9.3\times10^{-12}$\\
$\delta$--$L_{\rm LAT}$(All FSRQs)&41.67$\pm$0.49&5.35$\pm$0.41&40.09$\pm$0.86&6.71$\pm$0.72&40.97$\pm$0.61&5.95$\pm$0.51&0.89&$4.4\times10^{-16}$\\
$\delta$--$L_{\rm LAT}$(1H 0323+342)&42.36$\pm$0.09&4.25$\pm$0.17&42.24$\pm$0.13&4.44$\pm$0.24&42.30$\pm$0.10&4.34$\pm$0.19&0.98&$3.6\times10^{-3}$\\
$\delta$--$L_{\rm LAT}$(PMN J0948+002)&39.80$\pm$0.92&7.09$\pm$0.88&38.62$\pm$1.13&8.22$\pm$1.05&39.25$\pm$0.99&7.62$\pm$0.93&0.93&$3.0\times10^{-4}$\\
$\delta$--$L_{\rm LAT}$(All NLS1s)&41.67$\pm$0.28&5.23$\pm$0.31&41.24$\pm$0.44&5.71$\pm$0.44&41.47$\pm$0.35&5.46$\pm$0.36&0.96&$1.7\times10^{-9}$\\
$\delta$--$L_{\rm LAT}$(All)&41.51$\pm$0.22&5.48$\pm$0.20&40.98$\pm$0.35&5.96$\pm$0.31&41.25$\pm$0.28&5.71$\pm$0.24&0.96&$\sim0$\\
$\delta_{\rm F14}$--$\delta_{\rm LAT}$&0.84$\pm$0.02&0.49$\pm$0.03&0.65$\pm$0.03&0.72$\pm$0.04&0.75$\pm$0.02&0.60$\pm$0.03&0.82&$\sim0$\\
$\delta_{\rm H09}$--$\delta_{\rm LAT}$&0.93$\pm$0.09&0.29$\pm$0.08&-0.14$\pm$0.38&1.22$\pm$0.32&0.51$\pm$0.10&0.66$\pm$0.08&0.49&$8.8\times10^{-4}$\\
\noalign{\smallskip}\hline

\end{tabular}
\end{center}
\tablecomments{1\textwidth}{ ``All FSRQs" denote the 30 FSRQs (as described in Section 3) adding the 14 SEDs of 3C 279;``All NLS1s" denote the 17 SEDs of the five GeV NLS1s; ``All" denote all the SED data of both FSRQs and NLS1s.}

\end{table}

\section{The $\delta$--$L_{\rm c}$ Correlation in Different Sources}

The nine SED fits for NLS1 PMN J0948+0022 and five SED fits for NLS1 1H 0323+342 from Sun et al. (2015)\footnote{The slopes of linear fits for the two GeV NLS1s are slightly different from that reported in Sun et al. (2015) because the different linear fit methods are used.} are also presented in the $\delta$--$L_{\rm c}$ plane, as shown in Figure \ref{Lc-delta}(a). $L_{\rm c}$ is also tightly correlated with $\delta$ for the two GeV NLS1s, indicating that the variations of the luminosities are related with the variations of $\delta$ for an individual source in different stages. Considering the large error bars and small samples for the data of the two GeV NLS1s, the 3$\sigma$ confidence bands of the linear fits for the three sources are also separately given in Figure \ref{Lc-delta}(a). It can be found that the $\delta$--$L_{\rm c}$ relations for the three sources in different stages are consistent within 3$\sigma$ confidence bands. Therefore, the observed differences of $L_{\rm c}$ for the individual sources in different stages may be due to the different Doppler factors in different stages.

In order to examine whether the different sources share the same $\delta$--$L_{\rm c}$ relation with the individual sources, we compile a sample of 30 FSRQs\footnote{PKS 2142-758 in Zhang et al. (2015) is removed from our sample since the SED modeling results reported in  Zhang et al. (2015) are based on the SED reported in Dutka et al. (2012). However, we note that the SED of this source in almost the same temporal coverage in Dutka et al. (2013)  is dramatically different from that reported in Dutka et al. (2012). The data available in Dutka et al. (2012) would be only preliminary. We therefore do not use the SED modeling result of this source reported in Zhang et al. (2015) for analysis.} and five GeV NLS1s from our previous works (Zhang et al. 2014, 2015; Sun et al. 2015), in which their SEDs are measured simultaneously or quasi-simultaneously. The observed SEDs of FSRQs and GeV NLS1s have been systematically fitted with the one-zone leptonic models in our previous works (Zhang et al. 2014, 2015; Sun et al. 2015). Both $L_{\rm c}$ and $\delta$ are also obtained from our SED fits. We pick up only one SED for each FSRQs (Zhang et al. 2014, 2015), for which the same sources in the two papers the data in Zhang et al. (2015) are taken. Besides PMN J0948+0022 and 1H 0323+342 the other three GeV NLS1 galaxies in Sun et al. (2015) are also taken into account. The data of $L_{\rm c}$ and $\delta$ for the sample are reported in Table 3.

Figure \ref{Lc-delta}(b) shows $L_{\rm c}$ as a function of $\delta$ for these FSRQs and GeV NLS1s. It can be found that FSRQs and NLS1s form a clear sequence in the $\delta$--$L_{\rm c}$ plane. The linear fit yields $\log L_{\rm c}/{\rm erg\ s}^{-1}=(40.97\pm0.43)+(5.46\pm0.37)\log\delta$ with $r=0.93$ and $p=9.3\times10^{-14}$ for the 30 FSRQs, which is consistent within the error bars with the result of 3C 279 in different stages of $\log L_{\rm c}/{\rm erg\ s}^{-1}=(41.11\pm0.39)+(5.45\pm0.33)\log\delta$. It means that the different sources share a same relation with the individual sources in different stages. The linear fit to the combined sample of 30 FSRQs and 14 stages for 3C 279 gives $\log L_{\rm c}/{\rm erg\ s}^{-1}=(41.10\pm0.36)+(5.38\pm0.31)\log\delta$. The data adding of 14 stages for 3C 279 reduces the dispersion of this relation for 30 FSRQs. Since there are only five confirmed GeV NLS1 galaxies for which the data are available, the linear fit for the 17 SEDs of the five GeV NLS1s gives $\log L_{\rm c}/{\rm erg\ s}^{-1}=(41.09\pm0.33)+(5.39\pm0.36)\log\delta$ with $r=0.96$ and $p=5.5\times10^{-10}$. Hence the $\delta-L_{\rm c}$ relations are consistent within the error bars for the two kind of sources. The linear fit to the combined sample of all the FSRQs and GeV NLS1s yields $\log L_{\rm c}/{\rm erg\ s}^{-1}=(41.08\pm0.21)+(5.40\pm0.19)\log\delta$ with $r=0.97$ and $p\sim0$, as shown in Figure \ref{Lc-delta}(b).

\section{Physical Implications}

Note that $\delta$ is derived from the one-zone leptonic model fits for the broadband observed SEDs. Under the monochromatic approximation, $L_{\rm c}$ is given by
\begin{equation}
L_{\rm c}=\nu_{\rm c}L(\nu_{\rm c})=\frac{8\sigma_{\rm T}c}{9}U^{'}_{\rm ph}\gamma_{\rm b}^{3-p_1}R^3N_{0}\delta^4,
\end{equation}
where $\sigma_{\rm T}$ is the Thomson cross section, $c$ is the speed of light, $U^{'}_{\rm ph}$ is the energy density of photon field in the jet framework, $R$ is the radius of radiation region. If the EC process is dominated by IC/BLR, it is $U^{'}_{\rm ph}=(17/12)\Gamma^2U_{\rm BLR}$ (Ghisellini \& Madau 1996), where $U_{\rm BLR}$ is the energy density of BLR photon field at rest frame. Assuming $\delta=\Gamma$, we can obtain
\begin{equation}
L_{\rm c}=6\times10^{45}U_{\rm BLR,-2}P^{'}_{\rm e,56}\delta_{1}^6  \;\;\;\;\;\;\;\;\rm erg/s,
\end{equation}
where notation $Q_{\rm n}=Q/10^{\rm n}$, $P^{'}_{\rm e}=\frac{4}{3}\pi R^3N_{0}\gamma_{\rm b}^{3-p_1}$ could be a representative of the intrinsic powers of the radiation electrons for these sources.

\begin{figure*} 
\centering
 \includegraphics[width=\textwidth, angle=0]{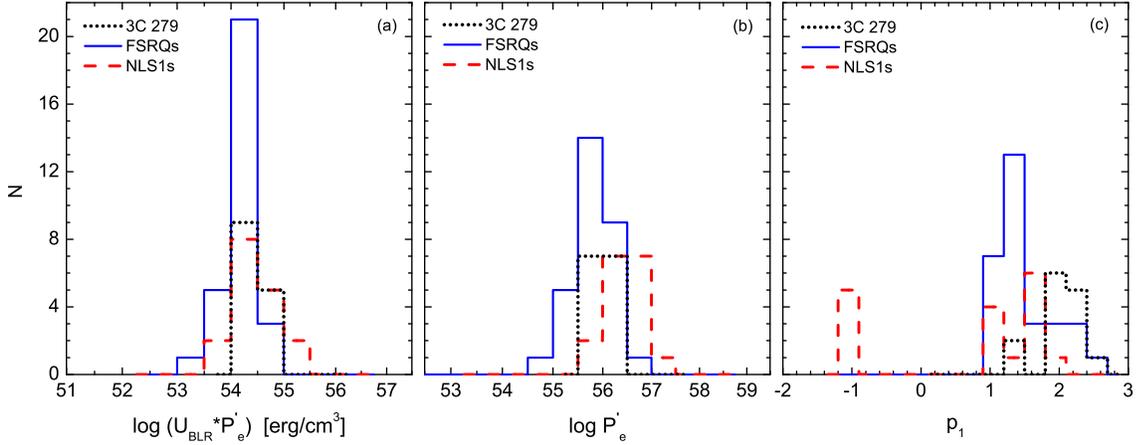}
 \caption{Distributions of $U_{\rm BLR}*P_{\rm e}^{'}$, $P_{\rm e}^{'}$, and $p_1$, where $P_{\rm e}^{'}$ is a representative of the intrinsic powers of the radiation electrons for these sources, $U_{\rm BLR}$ is the energy density of BLR at the rest framework, and $p_1$ is the spectral index of electron distribution.}
\label{coeffecient}
\end{figure*}

The analysis in Section 3 shows that FSRQs and GeV NLS1s share the same $\delta$--$L_{\rm c}$ relation, which may suggest a unification picture of the two kinds of sources, even an individual source in different stages. The linear fits yield $L_{\rm c}\propto\delta^{\sim5.4}$ for both FSRQs and GeV NLS1s, being roughly consistent with the model prediction as given in Equation (2). And thus the tight $\delta$--$L_{\rm c}$ correlation of both FSRQs and GeV NLS1s indicates that the values of $U_{\rm BLR}P^{'}_{e}$ are almost universal among sources and among different stages of these sources. As shown in Figure \ref{coeffecient}(a), it is indeed mostly distributed within 10$^{54}$--10$^{55}$ erg cm$^{-3}$ and both FSRQs and GeV NLS1s occupy the same region. This result indicates that $P_{\rm e}^{'}$ should be inverse proportional to $U_{\rm BLR}$. This is reasonable since a BLR with the larger $U_{\rm BLR}$ should be more effective to cool the electrons, resulting in a smaller $\gamma_{\rm b}$, and then a lower $P_{\rm e}^{'}$. The distributions of $P_{\rm e}^{'}$ of two kinds of sources are also presented in Figure \ref{coeffecient}(b). The $P_{\rm e}^{'}$ values for most of FSRQs and 3C 279 in different stages cluster at 10$^{55.5}$--10$^{56.5}$, implying that the acceleration energies of electrons in different stages and in different FSRQs are similar. Ones can also observe that the typical $P_{\rm e}^{'}$ value of FSRQs is smaller than that of GeV NLS1s. $U_{\rm BLR}$ in our previous SED modeling is taken as a constant or is calculated with the observed fluxes of emission lines (Zhang et al. 2014; 2015; Sun et al. 2015). The values of $U_{\rm BLR}$ are very close, but, on average, the derived values of $U_{\rm BLR}$ for FSRQs are slightly larger than that for GeV NLS1s. Therefore, the universal $U_{\rm BLR}P_{\rm e}^{'}$ value should be due to the EC cooling effect.

As reported in Sun et al. (2015), the jet radiation mechanisms and the circumnuclear environments for both FSRQs and NLS1s are similar. The universal value of $U_{\rm BLR}P^{'}_{e}$ among sources and among different stages of these sources may further suggest that their particle acceleration is also similar. To further investigate this issue, we show the distributions of $p_1$ for both FSRQs and NLS1s in Figure \ref{coeffecient}(c). It is found that the $p_1$ values for the FSRQs are in the range from 1 to 2.64, and they range from -1 to 2 for the GeV NLS1s. Most of them are clustered in $1\sim 2$ for the two kinds of sources. Note that the $p_1$ value expected from the first-order Fermi acceleration via relativistic shocks is larger than 2 (e.g., Kirk et al. 2000; Achterberg et al. 2001; Virtanen \& Vainio 2005). Therefore, the particle acceleration mechanism in these jets may not be dominated by the relativistic shocks. Magnetic reconnection may be the effective process of energy conversion and particle acceleration for jets in FSRQs and GeV NLS1s, which can produce a flatter power-law particle spectrum (Guo et al. 2015). This is also consistent with the moderately magnetized jets in both FSRQs and NLS1s (Zhang et al. 2013a; Zhang et al. 2015; Sun et al. 2015). However, we cannot also rule out the first-order Fermi acceleration mechanism since the derived indices of electron distribution for some sources or some stages of individual sources are consistent with the prediction of first-order Fermi acceleration mechanism as shown in Figure \ref{coeffecient}(c). Yan et al. (2016) also reported that the very hard electron spectrum can be produced using a time-dependent emission model in the fast cooling regime with KN effect. In addition, the stochastic acceleration scenario may be also an important particle acceleration mechanism in blazar jets (Virtanen \& Vainio 2005; Tramacere et al. 2011; Yan et al. 2013; Chen 2014). Recently, Petropoulou et al. (2016) suggested that the broadband SEDs of blazars can be produced by the different relativistic plasmoids (different $\delta$) with the same particle acceleration and radiation mechanisms. A universal $\delta$--$L_{\rm c}$ relation between FSRQs and GeV NLS1s in different stages may further evidence that the particle acceleration and radiation mechanisms in two kinds of sources are similar.

\section{Derivation of $\delta$ Values for FSRQs in 3FGL}

As discussed above, the observed differences of $L_{\rm c}$ in different stages and different sources would be governed by the Doppler boosting effect. Note that the peaks of EC/BLR bumps of FSRQs and NLS1s are usually in the \emph{Fermi}/LAT band. Therefore, we replace $L_{\rm c}$ (the peak luminosity of EC bump) with the observed luminosity in the LAT band ($L_{\rm LAT}$) and examine the $\delta$--$L_{\rm LAT}$ relation, where $L_{\rm LAT}$ is calculated with the energy fluxes from 100 MeV to 100 GeV. The $L_{\rm LAT}$ values of the corresponding SEDs for 3C 279, 30 FSRQs, and five GeV NLS1s are also given in Tables 1, 3. We plot $L_{\rm LAT}$ against $\delta$ in Figures \ref{Lc-delta}(c) and fit the $\delta$--$L_{\rm LAT}$ relations of sub-classes separately, which are given in Table 2. The bisector of the two OLS lines of linear regression fits gives $\log L_{\rm LAT}/{\rm erg\ s}^{-1}=(40.97\pm0.61)+(5.95\pm0.51)\log\delta$ with $r=0.89$ and $p=4.4\times10^{-16}$ for the 44 SEDs of FSRQs and $\log L_{\rm LAT}/{\rm erg\ s}^{-1}=(41.47\pm0.35)+(5.46\pm0.36)\log\delta$ with $r=0.96$ and $p=1.7\times10^{-9}$ for the 17 SEDs of NLS1s, respectively, which are consistent each other and also consistent with their $\delta$--$L_{\rm c}$ relation within the errors. Combined FSRQ and NLS1 data, the bisector of the two OLS lines of linear regression fits yields $\log L_{\rm LAT}/{\rm erg\ s}^{-1}=(41.25\pm0.28)+(5.71\pm0.24)\log\delta$ with $r=0.96$ and $p\sim0$, which is also consistent with their $\delta$--$L_{\rm c}$ relation. The Pearson correlation analysis shows that $L_{\rm c}$ is strongly correlated with the corresponding $L_{\rm LAT}$ with $r\sim0.98$ and $p\ll10^{-4}$. The distribution of $L_{\rm LAT}$/$L_{\rm c}$ clusters at $2\sim5$. These factors indicate that $L_{\rm LAT}$ is a good proxy of $L_{\rm c}$ and thus would be a good empirical indicator of $\delta$. Hence we can estimate the Doppler factors of FSRQs in 3FGL with the $\delta$--$L_{\rm LAT}$ relation.

\begin{table}
\begin{center}
\caption{The Data for the 30 FSRQs and Five GeV NLS1s in Our Sample}
\label{tab:2}
\setlength{\tabcolsep}{4pt}
\small
	\begin{tabular}{lccclccc}
	\hline\noalign{\smallskip}
	Sources  & $\delta$ & $\log L_{\rm c}$ & $\log L_{\rm LAT}$ & Sources  & $\delta$ & $\log L_{\rm c}$ & $\log L_{\rm LAT}$ \\
	 & &  [erg s$^{-1}$] &   [erg s$^{-1}$] & & &  [erg s$^{-1}$] &   [erg s$^{-1}$]  \\
	\hline\noalign{\smallskip}

	3C 273&7.4$\pm$0.9&45.95$\pm$0.30&46.55&3C 279&12.0$\pm$0.5&46.92$\pm$0.05&47.47\\
	3C 454.3&15.6$\pm$0.6&47.68$\pm$0.12&48.25&PKS 1454-354&20.2$\pm$1.8&48.13$\pm$0.25&48.66\\
	PKS 0208$-$512&15.8$\pm$0.7&47.41$\pm$0.15&47.89&PKS1502+106&27.0$\pm$2.3&48.60$\pm$0.22&49.31\\
	PKS 0420$-$01&12.8$\pm$0.7&47.06$\pm$0.13&47.59&B2 1520+31&20.8$\pm$1.6&47.95$\pm$0.25&48.47\\
	PKS 0528+134&17.4$\pm$0.9&48.38$\pm$0.14&48.77&4C 66.20 &12.2$\pm$1.2&46.82$\pm$0.12&47.54\\
	B3 0650+453&14.1$\pm$1.0&47.40$\pm$0.12&47.81&PKS 2325+093&17.6$\pm$1.6&47.98$\pm$0.17&48.43\\
	PKS 0727$-$11&20.6$\pm$1.2&48.19$\pm$0.15&48.74&1H 0323+342 (1)&2.8$\pm$0.6&43.83$\pm$0.30&44.28\\
	PKS 1127$-$145&13.1$\pm$0.8&47.25$\pm$0.16&47.45&1H 0323+342 (2)&3.6$\pm$1.3&44.24$\pm$0.40&44.65\\
	1Jy 1308+326&12.6$\pm$0.9&47.33$\pm$0.15&47.96&1H 0323+342 (3)&4.9$\pm$0.8&44.73$\pm$0.25&45.16\\
	PKS 1508$-$055&17.0$\pm$1.1&47.18$\pm$0.15&47.56&1H 0323+342 (4)&4.5$\pm$0.6&44.88$\pm$0.20&45.31\\
	PKS 1510$-$089&11.0$\pm$0.5&46.79$\pm$0.14&47.31&1H 0323+342 (5)&6.2$\pm$0.6&45.25$\pm$0.20&45.74\\
	TXS 1846+322&13.1$\pm$0.6&46.91$\pm$0.14&47.36&PMN J0948+0022 (1)&11.0$\pm$1.4&46.50$\pm$0.38&47.02\\
	PKS 2123$-$463&17.9$\pm$0.6&48.10$\pm$0.10&48.71&PMN J0948+0022 (2)&10.8$\pm$1.3&46.40$\pm$0.30&47.00\\
	TXS 2141+175&10.3$\pm$0.6&46.23$\pm$0.15&46.12&PMN J0948+0022 (3)&8.6$\pm$1.3&46.10$\pm$0.40&46.57\\
	PKS 2144+092&14.3$\pm$1.0&47.37$\pm$0.15&47.66&PMN J0948+0022 (4)&11.1$\pm$1.0&46.48$\pm$0.32&46.97\\
	PKS 2204$-$54&14.4$\pm$0.9&47.24$\pm$0.14&47.55&PMN J0948+0022 (5)&11.6$\pm$0.8&46.77$\pm$0.20&47.16\\
	PKS 2345$-$1555&13.8$\pm$1.0&46.80$\pm$0.18&47.15&PMN J0948+0022 (6)&9.5$\pm$0.5&46.49$\pm$0.20&46.90\\
	S4 0133+47&13.1$\pm$1.2&46.94$\pm$0.13&47.71&PMN J0948+0022 (7)&13.5$\pm$1.1&47.70$\pm$0.35&48.07\\
	PKS 0227$-$369&17.8$\pm$1.0&48.08$\pm$0.16&48.59&PMN J0948+0022 (8)&13.7$\pm$1.8&47.50$\pm$0.37&47.92\\
	4C 28.07&14.6$\pm$1.1&47.30$\pm$0.14&47.78&PMN J0948+0022 (9)&11.4$\pm$2.2&46.90$\pm$0.73&47.41\\
	PKS 0347$-$211&26.2$\pm$1.5&48.42$\pm$0.13&49.00&SBS 0846+513&7.4$\pm$0.8&46.03$\pm$0.18&46.61\\
	PKS 0454$-$234&20.0$\pm$1.9&47.77$\pm$0.09&48.38&PKS 1502+036&9.5$\pm$0.8&45.79$\pm$0.20&46.10\\
	S4 0917+44&18.2$\pm$1.3&47.86$\pm$0.13&48.44&PKS 2004$-$447&6.4$\pm$0.5&44.90$\pm$0.20&45.18\\
	4C 29.45&11.6$\pm$1.0&46.72$\pm$0.22&47.26&&&\\

	\noalign{\smallskip}\hline
\end{tabular}
\end{center}
\end{table}

\begin{figure*}
\centering
\includegraphics[angle=0,scale=0.23]{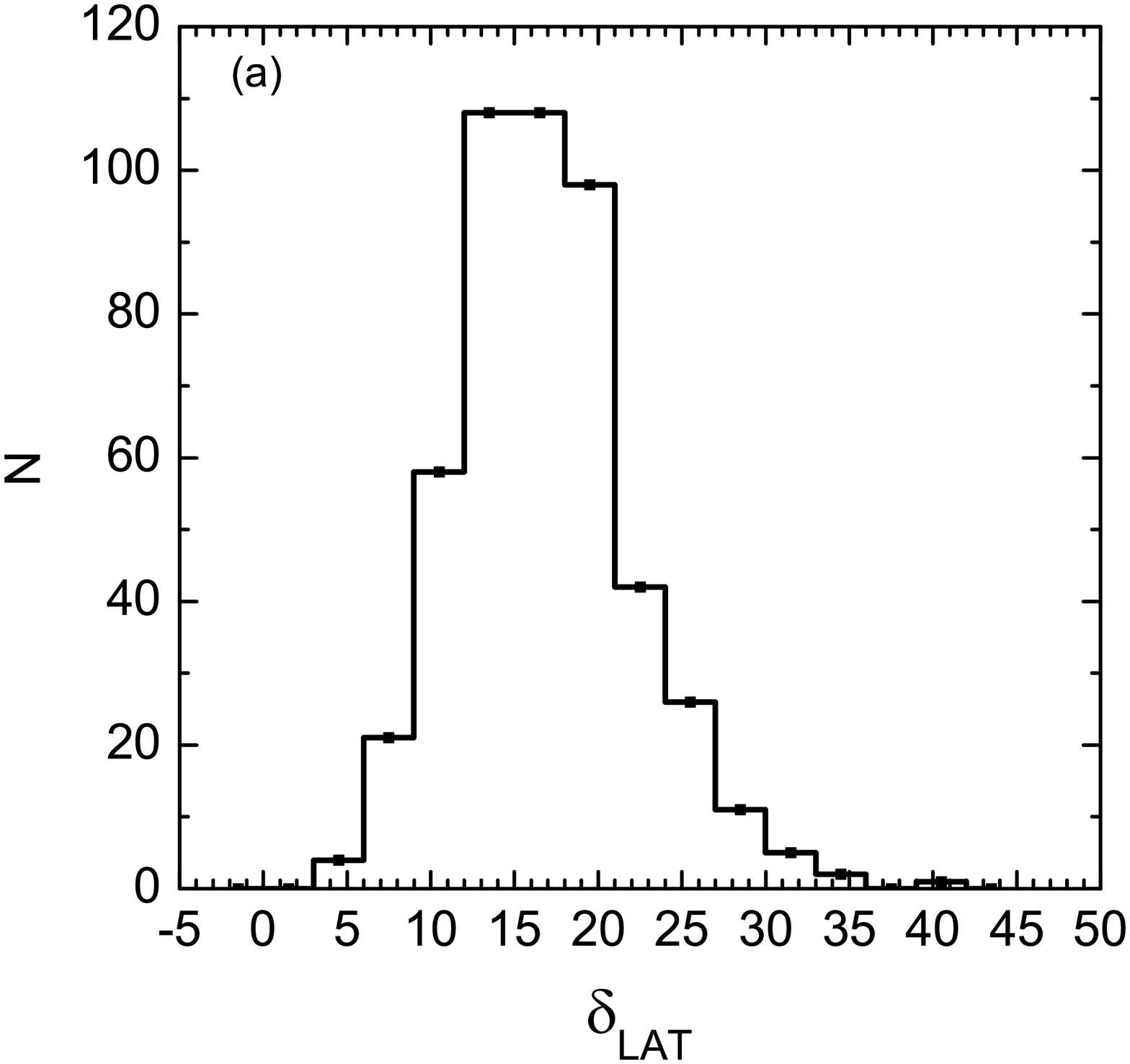}
\includegraphics[angle=0,scale=0.23]{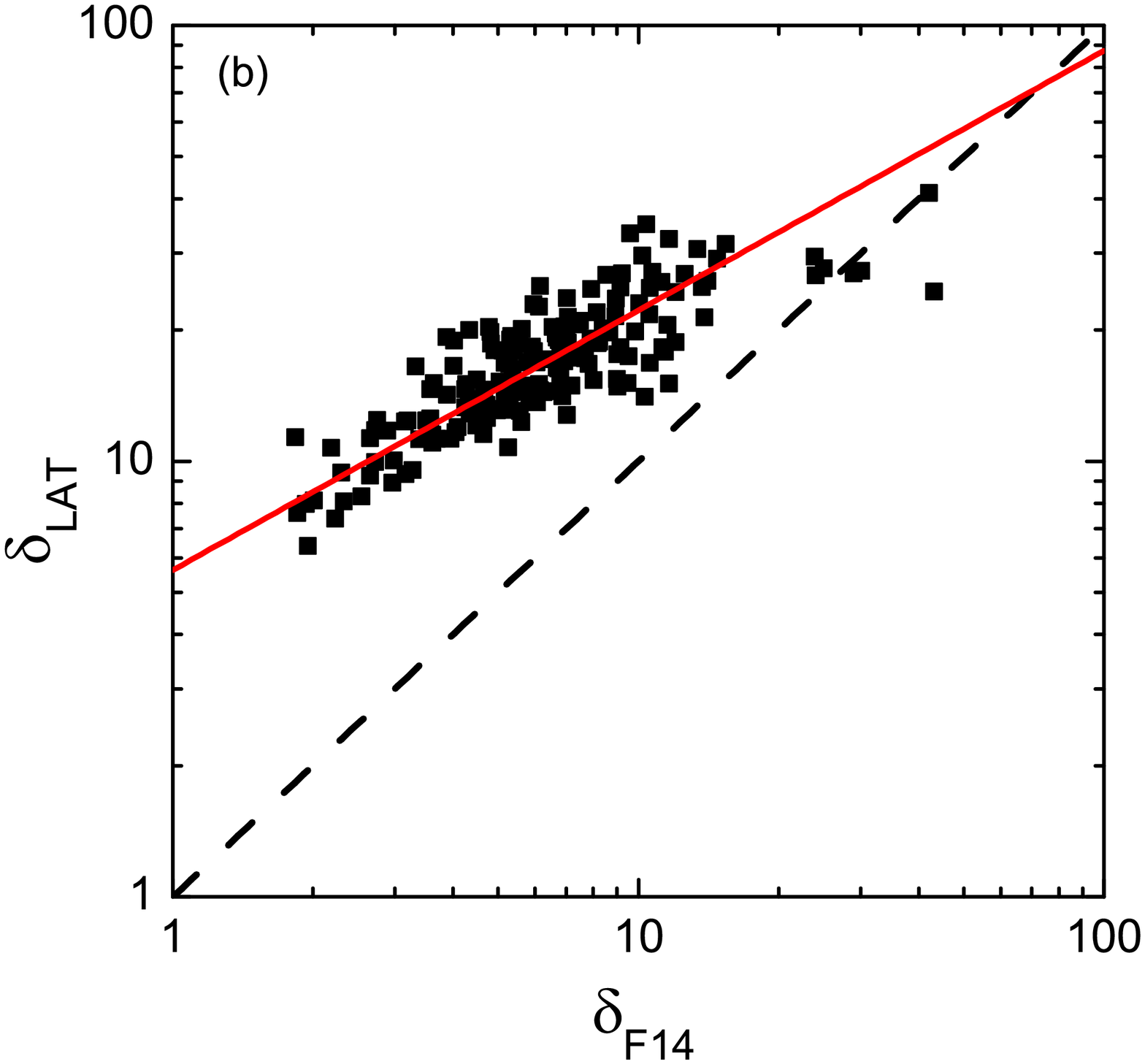}
\includegraphics[angle=0,scale=0.23]{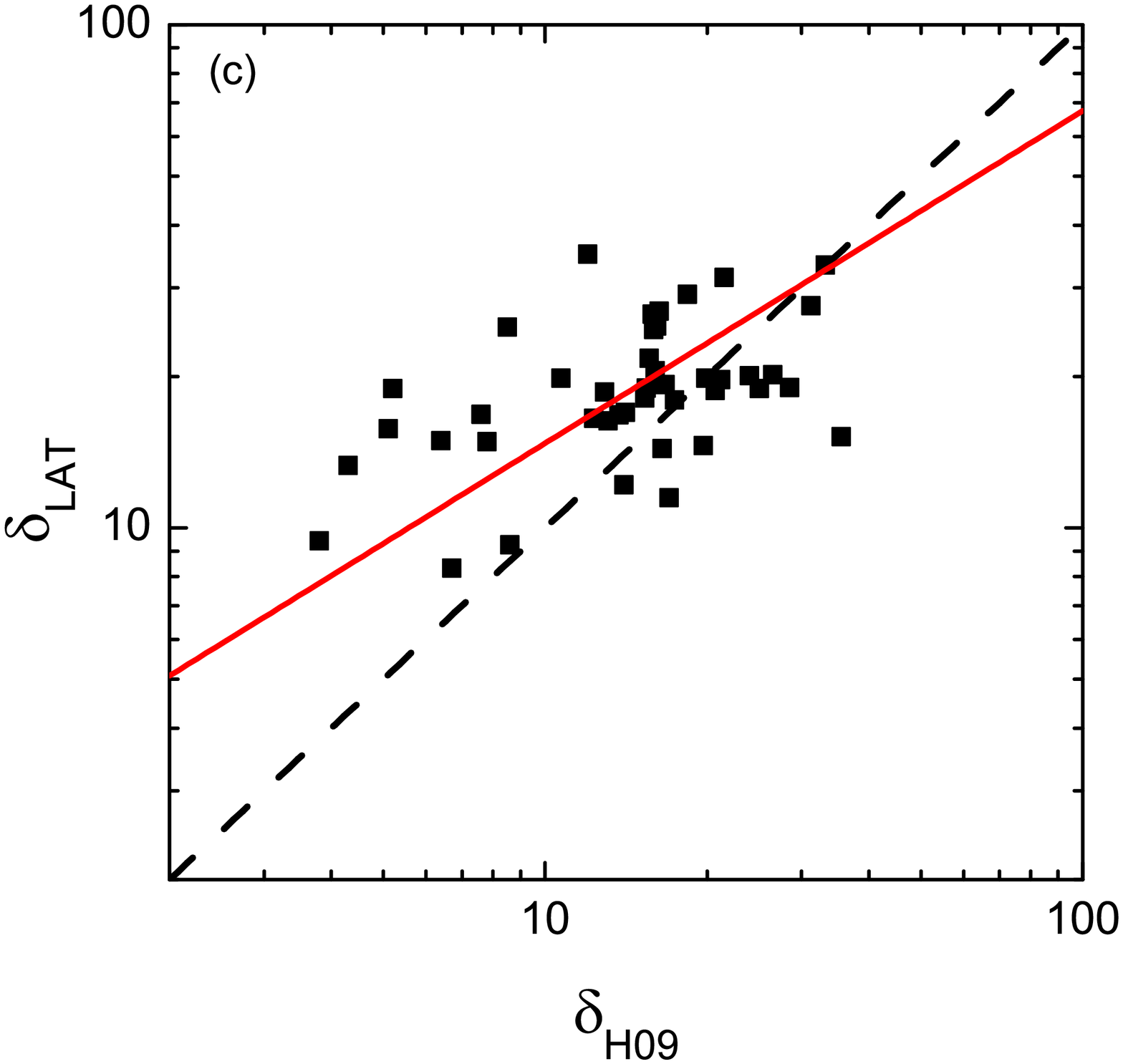}
\caption{\emph{Panel (a)}---Distribution of the derived Doppler factors by the $\delta$--$L_{\rm LAT}$ relation for 484 FSRQs with confirmed redshift in 3FGL. \emph{Panel (b)} and \emph{(c)}---Comparisons of the derived Doppler factors by $L_{\rm LAT}$ with that given in Fan et al.(2014) for 179 FSRQs (\emph{Panel (b)}) and in Hovatta et al.(2009) for 43 FSRQs (\emph{Panel (c)}), respectively. The black dashed lines are the equality lines and the red solid lines are the bisectors of the two OLS lines of linear regression fits with the parameters given in Table 2. }
\label{delta}
\end{figure*}

There are 484 FSRQs with confirmed redshift in 3FGL (Acero et al. 2015). Their $\delta_{\rm LAT}$ values are calculated using the $\delta$--$L_{\rm LAT}$ relation of the combined FSRQ and NLS1 sample with the available $L_{\rm LAT}$, i.e., $\log \delta=(-7.23\pm0.35)+(0.18\pm0.01)\log L_{\rm LAT}/{\rm erg\ s}^{-1}$. The $\delta_{\rm LAT}$ distribution of these FSRQs is shown in Figure \ref{delta}(a). The values of $\delta_{\rm LAT}$ range from 3 to 41 with a median of 16. With the data of the X-ray observations and the data of Second \emph{Fermi}/LAT Source Catalog, Fan et al. (2014) estimated the lower limits of Doppler factors ($\delta_{\rm F14}$) for the sources with the gamma-ray transparency condition of pair-production absorption (Mattox et al. 1993). There are 179 FSRQs that are included in the 484 FSRQs of 3FGL, and the comparison between $\delta_{\rm LAT}$ and $\delta_{\rm F14}$ for the 179 FSRQs\footnote{Since the data in the Second \emph{Fermi}/LAT Source Catalog represent an average state of the sources, the values of Doppler factors derived with the variability timescale of 1 day in Fan et al. (2014) are taken.} is presented in Figure \ref{delta}(b). It is found that they are strongly correlated with the Pearson correlation coefficient of $r=0.82$ and chance probability of $p\sim0$. The bisector of the two OLS lines of linear regression fits yields $\log \delta_{\rm LAT}=(0.75\pm0.02)+(0.60\pm0.03)\log\delta_{\rm F14}$, which is also shown in Figure \ref{delta}(b). Except for three sources\footnote{Each of them with high redshift has high flux and flat spectrum at X-ray band, which would result in the overestimates for the lower limits of their Doppler factors.}, $\delta_{\rm F14}$ are smaller than $\delta_{\rm LAT}$ for all the other FSRQs, which is reasonable because the values of Doppler factors given in Fan et al. (2014) are the lower limits of their Doppler factors.

Using the variability brightness temperatures of the fastest flares in the radio band, Hovatta et al. (2009) calculated the Doppler factors ($\delta_{\rm H09}$) of 87 AGNs. There are 43 FSRQs that are also included in the 484 FSRQs of 3FGL. We also compare $\delta_{\rm H09}$ with $\delta_{\rm LAT}$ for the 43 FSRQs, as given in Figure \ref{delta}(c). Most of the data points are above the equality line ($\delta_{\rm LAT} >\delta_{\rm H09}$), which is reasonable since $\delta_{\rm H09}$ is also a lower limit. A weak tentative correlation is observed with the Pearson correlation coefficient of $r=0.49$ and chance probability of $p=8.8\times10^{-4}$. The bisector of the two OLS lines of linear regression fits is also presented in Figure \ref{delta}(c). The larger scatter in Figure \ref{delta}(c) may be due to that the non-simultaneous observation data are used to calculate the values of Doppler factors. As reported in Hovatta et al. (2009), using the different flare data would yield different values of Doppler factors. Even for an individual source, the different observation luminosities at different time may be corresponding to the different Doppler factors (Zhang et al. 2013c; Sun et al. 2015). We test whether the two distributions of Doppler factors for the 43 FSRQs show any statistical difference with the Kolmogorov$-$Smirnov test (K--S test), which yields a chance probability ($p_{\rm KS}$). A K--S test probability larger than 0.1 would strongly suggest no statistical difference between the two distributions. We obtain $p_{\rm KS}=0.02$, indicating that the distribution of $\delta_{\rm LAT}$ is marginally consistent with the distribution of $\delta_{\rm H09}$ for the 43 FSRQs. These results suggest that the derived values of $\delta$ with the $\delta$--$L_{\rm LAT}$ relation are statistically consistent with the values calculated by other methods.

\section{Summary and Conclusions}

By modeling the broadband SEDs of a typical FSRQ 3C 279 and two GeV NLS1s PMN J0948+0022 and 1H 0323+342 in different stages with a one-zone leptonic model, we found a correlation between the Doppler factor ($\delta$) and EC peak luminosity ($L_{\rm c}$). We then compiled a sample of 30 FSRQs and 5 GeV NLS1 galaxies and found that the $\delta$--$L_{\rm c}$ correlation holds well. The main results are summarized as follows:

\begin{itemize}
\item $L_{\rm c}$ is strongly correlated with $\delta$ for both FSRQs and GeV NLS1s, and the two kinds of AGNs form a clear sequence in the $\delta$--$L_{\rm c}$ plane, which may imply a unified picture of the particle acceleration and cooling mechanisms in the comoving frame for the two kinds of sources. Therefore, the observed differences of $L_{\rm c}$ in different stages and different sources may be essentially due to their different Doppler factors.

\item Replacing $L_{\rm c}$ with the observed luminosity in the \emph{Fermi}/LAT band ($L_{\rm LAT}$), this correlation holds. The linear fitting result of the $\delta$--$L_{\rm LAT}$ relation is well consistent with the $\delta$--$L_{\rm c}$ relation within the errors. $L_{\rm LAT}$ may serve as an empirical indicator of $\delta$.

\item We estimated the $\delta_{\rm LAT}$ values with $L_{\rm LAT}$ for 484 FSRQs in 3FGL and they range from 3 to 41, with a median of 16. The derived values of $\delta$ are statistically consistent with the values calculated by other methods.

\end{itemize}

\begin{acknowledgements}
We appreciate helpful discussion with Bing Zhang. This work is supported by the National Basic Research Program (973 Programme) of China (grant 2014CB845800), the National Natural Science Foundation of China (grants 11573034, 11533003, 11373036, 11133002), the Strategic Priority Research Program ``The Emergence of Cosmological Structures" of the Chinese Academy of Sciences (grant XDB09000000), the Guangxi Science Foundation (2013GXNSFFA019001).
\end{acknowledgements}

\label{lastpage}

\end{document}